\documentclass[twocolumn,pra,showpacs,preprintnumbers,amsmath,amssymb]{revtex4}

\usepackage{graphicx}
\usepackage{dcolumn}

\def\be{\begin{eqnarray}}
\def\ee{\end{eqnarray}}
\def\bee{\begin{eqnarray*}}
\def\eee{\end{eqnarray*}}
\def\bra{\langle}
\def\ket{\rangle}
\def\kb{ \ket \bra }

   \def\tr{\hbox{Tr} \,}

\def\bmx{\begin{pmatrix}}
\def\emx{\end{pmatrix}}

\def\vrp{\varepsilon}

\begin{document}


\title{Exactly soluble model of resonant energy transfer between molecules}

\author{C. King$^1$, B. Barbiellini$^2$,  D. Moser$^{1,2}$ and V. Renugopalakrishnan$^{3,4}$}
\affiliation{%
$^{1}$Department of Mathematics, Northeastern University,
Boston, MA 02115 USA\\ 
$^{2}$Department of Physics, Northeastern University,
Boston, MA 02115 USA\\ 
$^{3}$Childrens Hospital, Harvard Medical School, Boston, MA 02115, USA \\
$^{4}$Department of Chemistry and Chemical Biology, Northeastern University, Boston, MA 02115 USA.
}%

\date{\today}

\begin{abstract}
F\"orster's theory of resonant energy transfer (FRET) 
predicts the strength and range of exciton
transport between separated molecules. We introduce an exactly soluble  
model for FRET which reproduces F\"orster's results as well 
as incorporating quantum coherence effects. As an application 
the model is used to analyze a system composed of quantum dots and 
the protein bacteriorhodopsin.
\end{abstract}


\maketitle

\section{Introduction}
It has long been known that near-field electrodynamics allows energy transfer without emission of real photons.
A striking example of this principle is provided by recent advances in wireless non radiative energy transfer 
\cite{KKMJFS2007}; another recent example is  Auger-mediated `sticking' \cite{mukerjee},  
whereby scattering states of positrons may transition to bound states in a metal,
by transferring energy to a valence electron which can then leave the surface of the metal.
Nowadays the phenomenon of non-radiative decay is of great scientific interest in many different fields of physics and chemistry \cite{bianconi}.
In particular, new paradigms for solar energy conversion make use of
nonradiative coupling for direct transfer of energy from the excitons created in the solar absorber to 
high mobility charge carriers \cite{BMG2010}, \cite{LSEFM2005}.
The mechanism for this transfer relies on the near-field resonance of electric dipoles,
and is generally known as F\"orster resonance energy transfer (FRET) 
\cite{Clegg,Scholes2003,For1946}.
In its simplest formulation, FRET \cite{foot1} is the quantum
version of a classical resonance phenomenon, whereby oscillating electric dipoles  exchange energy
through their mutual electric fields \cite{JPerrin}. 
Some studies in the quantum version have considered
the possibility of coherent interactions
between the dipoles 
\cite{RFRET0,RFRET1,RFRET2,RFRET3,RFRET4,sekatskii,LRNB2003,Hofmann2010}.
Such quantum coherence has been observed in the FMO complex, 
\cite{Engel2007}
and it has been suggested that this may partly explain the high efficiency of energy  transfer between
chromophores.
Interestingly, it has recently been demonstrated that
the classical dipole model reproduces the quantum coherence as well \cite{pre83,zimanyi}.
In this work, we revisit the analysis of the rate and efficiency of FRET, in the context of a donor and
an acceptor species with comparable electronic energy gaps.
In this situation FRET is evidenced by decreased natural fluorescence from
the donor, and enhanced fluorescence from the acceptor. The distance over
which FRET has been observed ranges from $1$ to  $10$ nm, with the strength
varying as the inverse sixth power of the separation.

\medskip
The fundamental mechanism underlying FRET is resonance between
excited electronic states in the donor and the acceptor molecules.
The excited electronic states have non-zero electric dipole moments, and the resulting
dipoles experience a Coulomb interaction. The energy exchange is complicated by 
the coupling of electronic states to vibronic molecular states, leading to a broadening of the
linewidths and a weakening of the resonant interaction. This vibronic coupling explains the
difference between early inaccurate calculations of FRET efficiency (by Perrin \cite{JPerrin} and others) 
which were based solely on dipole resonance, and the
later more successful calculations by F\"orster \cite{For1946} which included vibronic effects.

\medskip
In this paper we consider a simple model for FRET which incorporates 
both electronic and vibronic
effects. The model applies in situations 
where the donor molecule is rigid, with
weak coupling between its electronic and its vibronic states, 
while the acceptor has strong 
electronic-vibronic coupling in its excited state. In this situation the model
is exactly solvable and thus allows a comparison with the
perturbative formulas derived by F\"orster and others. In particular
we derive exact formulas for FRET efficiency and 
the F\"orster radius, and we compare these to the well-known F\"orster formulas.
Furthermore, the model
is fully quantum mechanical 
and predicts coherent oscillations between donor and
acceptor under strong FRET conditions.
The model contains a parameter which determines the strength of the 
electronic-vibronic coupling in the acceptor, and for weak coupling the model
reproduces the long-range interactions (up to $100$ nm) calculated by
Perrin. \cite{JPerrin} 

\medskip
As described above,
the model applies to a FRET system where a rigid donor species, with weak coupling between
electronic and vibronic states, interacts with an acceptor species where electronic and
vibronic states are strongly coupled. Thus the donor is modeled by a simple two-state system, corresponding
to its electronic ground and excited states. The donor's vibronic degrees of freedom are {\em frozen} and do
not appear in the model. As we discuss in more detail below, this kind of model can be
realized in practice with quantum dots (QDs). 
For the acceptor we again include only two electronic states,
corresponding to the ground and excited states, but in addition,
we include vibronic effects in the excited state. 
The excited band is described in detail
below. For the moment, we note that the electronic-vibronic 
coupling is derived from the Born-Oppenheimer approximation 
and assumes that the vibronic degrees of freedom are entrained to the electronic state.
Initially we assume that the vibronic degrees of freedom are also frozen in the acceptor ground state.
Later we indicate how nonzero temperature effects may be included by unfreezing these degrees
of freedom.

\medskip
The key step in the solution of our model is the reduction to a finite dimensional system
which exhibits the same efficiency for resonant energy transfer. The efficiency can be computed
exactly for this finite-dimensional model, and this provides our exact results for the full model.
In order to compare with F\"orster's formulas, we also compute the absorption coefficient for our model,
and we use this to evaluate the overlap integral which appears in the standard rate formulas.

\medskip
As an application we apply our model to the analysis of one particular system of this type which exhibits FRET, 
namely the pairing of a QD donor species 
with bacteriorhodopsin (bR) as the acceptor.
The QD is known to have sharp emission lines 
for fluorescence and, thus, is a good
candidate for the {\em rigid} donor molecule described above.
We find good agreement with other calculations of the 
FRET rate for this system.

\section{Description of the model}
We employ the Born-Oppenheimer approximation
and assume that the electronic state of the molecule determines its overall
character, so that the vibronic state is entrained to the electronic state. Thus the state space is a direct sum
${\cal H}_0 \oplus {\cal H}_1 \oplus \cdots$ with one factor for each electronic state.
The electronic configuration determines an effective Hamiltonian for the vibronic degrees of freedom,
and each space ${\cal H}_{k}$ is spanned by these vibronic states attached to the corresponding ground state.
The spaces ${\cal H}_{k}$ may be discrete or continuous, depending on the structure of the
vibronic states.
For simplicity we include only two electronic states, the ground state and the excited state,
thus the state space is ${\cal H}_{gr} \oplus {\cal H}_{exc}$.

\subsection{The ground subspace ${\cal H}_{gr}$}
We assume a non-degenerate electronic ground state $| \psi_{gr} \ket$, and we also assume that the vibronic modes are frozen, so
${\cal H}_{gr}$ is one-dimensional. For the acceptor this assumes zero temperature.
Later we extend to nonzero temperatures by 
including vibronic  ground states.

\subsection{The excited subspace ${\cal H}_{exc}$}
Again, we assume a non-degenerate electronic excited state $| \psi_{exc} \ket$.
For the donor, the vibronic modes are frozen, so that ${\cal H}_{exc}$
is one-dimensional. However, for the acceptor,
the electronic state determines an effective Hamiltonian for the vibronic 
states, which are labeled by their energy eigenvalues $\vrp$. 
For this subspace we assume that the vibronic states form a continuous 
band with a uniform density of states, with eigenvalues extending 
from $- \infty$ to
$+ \infty$. Under this assumption we ignore any edge effects in the band.
Thus the space ${\cal H}_{exc}$ is isomorphic to the one-particle
Hilbert space $L^2(\mathbb{R})$, and we represent a state as a square integrable
function $\phi_{exc}(\vrp)$ where
\be
\int_{-\infty}^{\infty} |\phi_{exc}(\vrp)|^2 d \vrp = 1~.
\ee
The time evolution is $\phi_{exc}(\vrp) \rightarrow e^{- i \vrp t} \phi_{exc}(\vrp)$.
For convenience we denote the Hamiltonian in this basis as $h$, so that
\be
(h \phi_{exc})(\vrp) = \vrp \, \phi_{exc}(\vrp)~.
\ee

\subsection{Excitons}
As long as the electronic state does not change, the dynamics
of the vibronic state is completely determined by the fixed effective Hamiltonian corresponding to
this electronic configuration (here we are neglecting any feedback reaction from the vibronic
modes on the electronic modes). However, when the molecule undergoes
an electronic transition, for example by photon absorption, the effective Hamiltonian for the vibronic
states immediately changes. This sudden change creates an excited vibronic state, as the previously stationary vibronic state
becomes a superposition of energy eigenstates states of the new Hamiltonian.
We call this vibronic state an {\it exciton}. The exciton behaves like a delocalized one-particle
state. In our model, exciton states will arise
only in the excited band of the acceptor, due to a transition from the ground state.
We assume an average energy $E_3$
for these exciton states.

\subsection{Transitions}
Turning now to transitions, we consider only radiative 
interactions which act 
solely on the electronic state. Thus transitions of the vibronic state occur as a consequence of the
change of the effective Hamiltonian due to the electronic transition.
The electronic matrix element due to the interaction $V$ is
\be\label{matr-el-1}
\bra \psi_{exc} | V | \psi_{gr} \ket~. 
\ee
We find an explicit form for this matrix element for the situations of interest, namely,
direct photon absorption and resonant excitation through the Coulomb interaction.
In order to determine the vibronic matrix element, 
we follow Jortner \cite{Jortner}
and proceed by analogy with the derivation of the lineshape of a resonance.
Recall that a resonance is a perturbation of an embedded eigenvalue in continuous
spectrum. The perturbation causes the eigenvalue 
to {\em dissolve}, accompanied by
the emission of a one-particle state in the continuous spectrum. 
In our model this one-particle state is the exciton. 
Thus the transition from a vibronic ground state is accompanied by the 
creation of an exciton, which is a normalized excited vibronic state.
The Breit-Wigner form for the lineshape
of the resonance is a Lorentzian, where the width corresponds to the
lifetime of the resonance, \cite{Jortner} and the center is the average energy.
We assume the same form for the exciton, so the wave function of the exciton
(in the diagonal energy representation) is the square root of a Lorentzian:
\be\label{Lorentz1}
f(\vrp) =  \sqrt{\frac{\gamma}{2 \pi}} \frac{e^{i \theta}}{(\vrp - E_3) + \frac{i}{2} \gamma}~.
\ee
Using a Lorentzian form implicitly assumes that the energy band extends from $- \infty$ to
$+ \infty$. We make this assumption, thus ignoring any edge effects in the band.
The width $\gamma$ depends on the particular system and determines the lifetime of the
exciton. The Lorentzian is centered at energy $E_3$, corresponding to the average exciton energy.
We also include a phase factor $e^{i \theta}$, which may depend on $\vrp$.
As we will see this phase factor is irrelevant to the calculation of the FRET efficiency.

\section{Dynamics of the model}
The FRET interaction between the donor and the acceptor causes an exchange
of energy as the excited state is transferred from one to the other. Although the
transfer becomes irreversible after some time, the initial interaction is
unitary and thus admits the possibility of oscillations between donor and acceptor.

\medskip
We use standard notation, with $D$ for donor ground state, $D^*$ for donor excited state,
$A$ for acceptor ground state, and $A^*$ for acceptor excited state.
Thus the coupled donor-acceptor system
is described by the collection of states
\be
| D A \ket, \, | D^* A \ket, \, | D A^* \ket, \, | D^* A^* \ket,
\ee
where both $| D A \ket $ and $ | D^* A \ket$ are single states, while 
$| D A^* \ket $ and $ | D^* A^* \ket$ 
contain the exciton subspace.

\subsection{The Hamiltonian}
The FRET interaction is dipole-dipole in lowest order, and so
its strength decays as the inverse third power of the
distance. The Hamiltonian is determined by the matrix element
of the Coulomb interaction $V_C$ between the electronic parts of the states $| D^* A \ket$ and $| D A^* \ket$.
Thus the electronic transition matrix element is
\be
\label{def:U}
U & = & \bra D A^* | V_C | D^* A \ket \\ \nonumber
& =& \frac{1}{R^3} \left( {\bf D}_D \cdot {\bf D}_A - \frac{3}{R^2} ({\bf D}_D \cdot {\bf R})
({\bf D}_A \cdot {\bf R}) \right)~,
\ee
where $R$ is the separation between the systems, and $D_D$ and $D_A$ are the transition dipole moments of the
donor and acceptor, respectively.
We use atomic units throughout, and we assume that
the dielectric constant is $1$.
We look in detail at specific models later, but for the moment 
we note that for typical systems
the dipole moment is about $10$ D, 
so at separations of around $5$ nm 
the interaction energy $U \sim 10^{-4}$ eV,
compared to the typical energy gaps between 
ground and excited states of $1 - 2$ eV.

\medskip
In the absence of other effects, we could analyze the dynamics of this coupled system
by restricting to the
subspace spanned by the states $| D^* A \ket$ and $| D A^* \ket$, and computing the time evolution
of the initial state $| D^* A \ket$ under the influence of the interaction $U$. The Hamiltonian is
\be
H = \bmx E_1 & \overline{U} \bra f | \cr U | f \ket & E_2 + h \emx~,
\ee
where $E_1$ and $E_2$ are the electronic energy gaps of the donor and acceptor, respectively;
$U$ is the interaction matrix element defined in Eq.~(\ref{def:U});
and $h$ is the diagonal 
energy operator of the continuous exciton band in the excited state. 
Also, $| f \ket$ denotes the
creation operator for the exciton as in Eq.~(\ref{Lorentz1}), 
and $\bra f |$ denotes the corresponding
annihilation operator.

\subsection{The master equation}
In our model we also include the effects of fluorescent decay from the
excited state to the ground state for both systems. We do this by introducing jump operators for the (irreversible)
fluorescent decays from excited to ground state 
and use a master equation to
compute the time evolution of the density matrix. So we are using the Markov
approximation for the coupling to the electromagnetic field which causes fluorescence.\cite{foot2}
In order to separate the outcomes from the two excited states
$| D^* A \ket$ and $| D A^* \ket$, 
we use two copies of the ground state to indicate which molecule has
decayed (since the jump operators are irreversible, there is no coupling from these ground 
states back to the excited states). Thus the system is represented by a density matrix $\rho$,
spanning the states $| D^* A \ket$, $| D A^* \ket$, $| \hat{D} A \ket$, 
and $| D \hat{A} \ket$, where
$| \hat{D} A \ket$ and $| D \hat{A} \ket$ are copies of the ground state, and
the two fluorescent decay channels are $| D^* A \ket \rightarrow | \hat{D} A \ket$ and
$| D A^* \ket \rightarrow | D \hat{A} \ket$. In this subspace the Hamiltonian is
\be
H_{DA} = \bmx  E_1 & \overline{U} \bra f | & 0 & 0 \cr U | f \ket & E_2 + 
h & 0 & 0 \cr 0 & 0 & 0 & 0 \cr 0 & 0 & 0 & 0  \emx~.
\ee
The effects of fluorescence are implemented by the jump operators:
\be
J_1 = \bmx  0&0&0&0 \cr 0&0&0&0 \cr 1&0&0&0 \cr 0&0&0&0 \emx, \quad
J_{2,i} = \bmx 0&0&0&0 \cr 0&0&0&0 \cr 0&0&0&0 \cr 0& \bra u_i | &0&0 \emx,
\ee
where $\{| u_i \ket\}$ form an orthonormal basis in the exciton space.
The master equation is
\be\label{eqn:mas1}
\frac{d \rho}{d t} &=& - i [H_{DA}, \rho] \\
\nonumber
&&+ \frac{\gamma_1}{2} \left(2 J_1 \rho J_1^{\dagger} - J_1^{\dagger} J_1 \rho - \rho J_1^{\dagger} J_1 \right) \\
\nonumber
&&+\sum_{i} \frac{\gamma_2}{2} \left(2 J_{2,i} \rho J_{2,i}^{\dagger} - J_{2,i}^{\dagger} J_{2,i} \rho
- \rho J_{2,i}^{\dagger} J_{2,i} \right)~,
\ee
where $\gamma_1,\gamma_2$ are the rates for fluorescence $| D^* A \ket \rightarrow | \hat{D} A \ket$
and $| D A^* \ket \rightarrow | D \hat{A} \ket$, respectively.

\subsection{Definition of the efficiency, F\"orster radius and FRET rate}
Using the master equation in Eq.~(\ref{eqn:mas1}), 
it is possible to compute the probability of exciton
transfer from the donor to the receiver 
and to find the efficiency of this process.
In the absence of FRET, the system ultimately ends up in 
state $| \hat{D} A \ket$.
Thus we define the efficiency of FRET to be the long-run probability that this does not happen; that is,
\be\label{def:E}
F = 1 - \lim_{t \rightarrow \infty} \bra  \hat{D} A  | \rho(t) |  \hat{D} A  \ket~.
\ee

\medskip
The F\"orster radius $R_0$ is then defined by the condition that at this separation the efficiency
reaches $50\%$. That is,
\be\label{def:R0}
R_0 = \max \{ R \,:\, F \ge 0.5 \}.
\ee

\medskip
The FRET rate $\gamma_{FRET}$ can also be computed from the efficiency, by comparing the
rates for fluorescence of the donor and FRET:
\be
\gamma_{FRET} = \gamma_1 \, \frac{F}{1 - F}~,
\ee
where $\gamma_1$ is the natural fluorescence rate for the donor. This follows from the relation
\be
F = \frac{\gamma_{FRET}}{\gamma_{FRET} + \gamma_{1}}~.
\ee

\section{Solution of the master equation}
We solve the master equation given in Eq.~(\ref{eqn:mas1})
with the initial condition $\rho(0) = | D^* A \kb D^* A |$, corresponding to the donor in its
excited state and the acceptor in the ground state. The solution of Eq.~(\ref{eqn:mas1})
takes the block diagonal form
\be\label{rho42}
\rho(t) = \bmx  \rho_{11}(t) & \rho_{12}(t) & 0 & 0 \cr \rho_{21}(t) & \rho_{22}(t) & 0 & 0 \cr
0 & 0 & \rho_{33}(t) & 0 \cr 0 & 0 & 0 & \rho_{44}(t)  \emx~.
\ee
The top left $2 \times 2$ block is
\be
\bmx \rho_{11}(t) & \rho_{12}(t)  \cr \rho_{21}(t) & \rho_{22}(t)  \emx 
= e^{- i B t} \, \bmx  1 & 0 \cr 0 & 0  \emx \, e^{i B^* t}~,
\ee
where $B$ is the non-Hermitian operator
acting on the space spanned by $| D^* A \ket$ and $| D A^*(\vrp) \ket$,
i.e.,
\be\label{def:B2}
B = \bmx E_1 - \frac{i}{2} \, \gamma_1 & \overline{U} \, \bra f | \cr
 U \, | f \ket &  E_2 + h - \frac{i}{2} \, \gamma_2  \emx,
\ee
and $| f \ket$ is the Lorentzian function in Eq.~(\ref{Lorentz1}).
The remaining diagonal entries in Eq.~(\ref{rho42}) are given by
\be\label{rho52}
\rho_{33}(t) &=& \gamma_1 \, \int_{0}^t  \rho_{11}(s) \, d s \\
\rho_{44}(t) &=& \gamma_2 \, \int_{0}^t \tr \rho_{22}(s) \, d s~.
\ee
In order to facilitate the notation, define the state
\be
| \psi_0 \ket = \bmx  1 \cr 0  \emx~.
\ee
It follows that the efficiency is given by
\be\label{sol-eff2}
F = 1- \rho_{33}(\infty) = 1 - \gamma_1 \, \int_{0}^{\infty} \left| \bra \psi_0 | e^{- i B s} | \psi_0 \ket \right|^2 \, d s.
\ee
Thus the calculation reduces to the problem of finding matrix elements of the operator $e^{- iB s}$. 
This is straightforward because $B$ is a rank $1$ perturbation of a diagonal operator.
The key step is the reduction to a related two-state system.
Namely, define the $2 \times 2$ matrix
\be\label{def:B3}
{\hat B} =   \bmx E_1  - \frac{i}{2} \, \gamma_1 & \overline{U} \cr
 U  &  E_2  + E_3 - \frac{i}{2} \, (\gamma + \gamma_2)  \emx.
\ee
It is shown in Appendix A that
\be
\bra \psi_0 | e^{- i B s} | \psi_0 \ket = \bra \psi_0 | e^{- i {\hat B} s} | \psi_0 \ket, 
\ee
and thus
\be\label{sol-eff3}
F = 1 - \gamma_1 \, \int_{0}^{\infty} \left| \bra \psi_0 | e^{- i {\hat B} s} | \psi_0 \ket \right|^2 \, d s.
\ee
Thus the effect of the exciton coupling in this model is the same as in
a two-state model with a second channel for
decay of the excited state to the ground state, at a rate 
which is the inverse lifetime of the exciton. 

\medskip
The efficiency $F$ in (\ref{sol-eff3}) can be evaluated by finding the eigenvectors and eigenvalues of ${\hat B}$.
The assumption that $\gamma_1, \gamma_2 > 0$
implies that ${\hat B}$ has two eigenvalues with negative imaginary parts.
These can be computed explicitly, and the initial state $| \psi_0 \ket$ can be written as a linear combination
of the eigenvectors. The result is
\be\label{computeF1}
F = \frac{ (1 + r) |U|^2}{(1 +r)^2 |U|^2 +  
4 r  \left[ \frac{\Gamma^2}{16} + (E_1  - E_2 - E_3)^2\right]},
\ee
where
\be
\Gamma = \gamma_1 + \gamma_2+ \gamma~,
\ee
and
\be
r = \frac{\gamma_1}{\gamma_2+\gamma}~.
\ee

\subsection{F\"orster radius and FRET rate}
From Eq.~(\ref{computeF1}) we compute the F\"orster radius and the FRET efficiency for this model.
Setting $F = 1/2$ we get the condition
\be
|U|^2 &=& \frac{4r}{1 - r^2} \\ \nonumber 
& & \left[ \bigg(\frac{\gamma_1+\gamma_2+\gamma}{4}\bigg)^2 + (E_1   - E_2 - E_3)^2\right]~.
\ee
The interaction $U$ is given by Eq.~(\ref{def:U}). Introducing an angular factor $\kappa$ this gives
\be
|U|^2 = \kappa^2 \frac{D_D^2 D_A^2}{R^6}
\ee
and, hence, the formula for the F\"orster radius
\be\label{R0-1}
R_0^6 = \frac{1 - r^2}{4 r} \frac{\kappa^2 D_D^2  D_A^2}{\frac{1}{16} \,
\bigg(\gamma_1+\gamma_2+\gamma\bigg)^2 + (E_1   - E_2 - E_3)^2}
\ee

\subsection{Exciton lifetime and scaling at resonance}
At resonance where the energies match we have $E_1   - E_2 - E_3 = 0$.
In this case we have
\be
F = \frac{\gamma_2 + \gamma}{\gamma_1 + \gamma_2 + \gamma} \,
\frac{|U|^2}{|U|^2 + \frac{\gamma_1(\gamma_2 + \gamma)}{4}}~.
\ee
Define the dimensionless parameter
\be
\eta = \frac{4 |U|^2}{\gamma_1(\gamma_2+\gamma)};
\ee
then at resonance we get
\be
F = \frac{\gamma_2 + \gamma}{\gamma_1 + \gamma_2 + \gamma} \,
\frac{\eta}{\eta+1}.
\ee
It is reasonable to expect that the inverse exciton lifetime $\gamma$ is much larger than the
fluorescence rates $\gamma_1,\gamma_2$. In this case the FRET efficiency at resonance
takes the simple form
\be
F = \frac{\eta}{\eta+1}, \quad \eta = \frac{4 |U|^2}{\gamma_1 \, \gamma}.
\ee
This same scaling relation has been found for  resonant energy transfer between
classical oscillators \cite{KKMJFS2007}, and seems to be  a general feature of this type of
phenomenon.

\subsection{Temperature dependence}
In order to incorporate non-zero temperature effects, we introduce vibronic ground states
for the acceptor, labeled $\{ | \phi_{gr}(\vrp_1) \ket \}$,
where $\vrp_1$ is the energy in the
ground state vibronic band.
These states are assumed non degenerate, and thus completely 
labeled by their energy eigenvalue
$\vrp_1$, with respect to the Hamiltonian determined by the electronic state $| \psi_{gr} \ket$. 
Thus a general state in the acceptor band will be
\be
| A  \ket = \sum_{\vrp_1}  c(\vrp_1) \,  | \phi_{gr}(\vrp_1) \ket,
\ee
with the normalization condition $\sum_{\vrp_1}  |c(\vrp_1)|^2 = 1$. 
The vibronic Hamiltonian is diagonal in this representation, and so the time evolution
of a state is
\be
| A  \ket \rightarrow \sum_{\vrp_1}  c(\vrp_1) \, e^{- i \vrp_1 t}  | \phi_{gr}(\vrp_1) \ket.
\ee
To include nonzero temperatures,
we assume a Boltzmann distribution for the initial equilibrium state; 
that is,
\be
\rho_{eq} = Z^{-1} \, \sum_{\vrp_1} e^{-\vrp_1/k_B T} \, | \phi_{gr}(\vrp_1) \kb \phi_{gr}(\vrp_1) |.
\ee
We then compute the thermal average of the efficiency over initial
vibronic energies. Thus the temperature-dependent efficiency is
\be\label{Boltz1}
F(T) = Z^{-1} \,\sum_{\vrp_1} e^{-\vrp_1/k_B T} \,
F(\vrp_1) \, d \vrp_1.
\ee
where $F(\vrp_1)$ is given by Eq.~(\ref{computeF1}) with $E_1$ replaced by the initial
energy $E_1 + \vrp_1$. By using this formula for $F(\vrp_1)$
we assume that initially the acceptor is in
a pure vibronic ground state $| \phi_{gr}(\vrp_1) \ket$, which  transitions to the exciton state
$ f(\vrp)$ due to the FRET interaction. Our main simplification is the following:
we assume that the reverse operation causes a transition from the exciton state back to
the {\em same initial vibronic state} $| \phi_{gr}(\vrp_1) \ket$. So under this reverse operation an exciton state
$\psi(\vrp)$ is mapped to  $\alpha \, | \phi_{gr}(\vrp_1) \ket$, where the
amplitude is $\alpha = \int \overline{f(\vrp)} \psi(\vrp_2) d \vrp$.
Thus the acceptor always returns to its initial ground state
(this assumption has also been for exact calculations of 
coherent exciton scattering \cite{vibron}).

\medskip
Carrying out the summation in Eq.~(\ref{Boltz1}) requires knowledge of the phonon
spectrum of the acceptor, and we do not pursue the question further here. However, we note that this effect of the
temperature is expected to be small because at room temperature $k_B T$ is significantly smaller than the energy scale 
given by $\gamma\sim 0.1$ eV.

\section{Comparison with standard FRET formulas}
The standard F\"orster formulas for FRET rate and F\"orster radius
involve the overlap integral between the normalized donor fluorescence spectrum
and the acceptor absorption spectrum. The formula for the 
F\"orster radius (in cm) is \cite{For1946}
\be\label{RF1}
R_F^6 = \frac{9000 \ln (10) \kappa^2 Q_D}{128 \pi^5  N_A} \,J,
\ee
where $Q_D$ is the donor quantum yield, $N_A$ is Avogadro's number,
and $J$ is the overlap integral (in cm$^3$dm$^3/$mol),
\be
J = \int_{0}^{\infty} \epsilon_A(\lambda) F_D(\lambda) \, \lambda^{4} \, d \lambda
\ee
(recall that we have assumed that the refractive index is $1$).
Here $F_D(\lambda)$ is the donor emission spectrum
[normalized so that $ \int_{0}^{\infty} F_D(\lambda)  \, d \lambda = 1$, where $\lambda$ is the wave length (in cm)], and
$\epsilon_A$ is the molar absorption coefficient in 
(cm$^{-1}$ dm$^3/$mol). Inserting values 
for the constants gives
\be
R_F^6 = 8.79 \times 10^{-25}  \, Q_D \, \kappa^2 \, J.
\ee
In our two-level model the donor is assumed to have a narrow
band fluorescence spectrum centered at energy $E_1$, so the overlap integral is essentially 
$\epsilon_A(\lambda)  \, \lambda^{4}$ 
evaluated at the wavenumber $\lambda = 2\pi c/E_1$. Thus the formula is
\be\label{RF3}
R_F^6 = 8.79 \times 10^{-25}  \, Q_D \, \kappa^2 \, \epsilon_A(2\pi c/E_1) \,
\, (2\pi c/E_1)^{-4}.
\ee

\medskip
The acceptor's molar absorption coefficient can be
evaluated from the standard absorption rate for the transition
from ground state to excited state, using Fermi's Golden Rule to compute the rate.
The details are carried out in Appendix B and the result 
(in cm$^{-1}$ dm$^3/$mol$=10$ m$^2$/mol) is
\be\label{accept-absorp1}
\frac{\epsilon_A \ln(10)}{N_A a_B^2} =
\frac{2 \pi}{\omega c}  \, \frac{10 E_2^2 D_A^2 \gamma}{(\omega  - E_2 - E_3)^2
+ \gamma^2/4},
\ee
where $a_B$ is the Bohr radius.
Using this expression for the absorption coefficient in Eq.~(\ref{RF1})
and setting $\omega = E_1$, we get (in a.u.)
\be\label{R0:stan}
R_F^6& = & 9 \, \kappa^2 Q_D \left(\frac{c}{E_1}\right)^3\,\\
\nonumber 
&&\frac{\gamma D_A^2}{(E_1  - E_2 - E_3)^2
+ \gamma^2/4} \, \left(\frac{E_2}{E_1}\right)^2~. \,
\ee

\medskip
Turning now to our Eq.~(\ref{R0-1}), we use
the Einstein A coefficient to relate the fluorescence rate and the dipole strength of the donor:
\be\label{Einstein}
\gamma_1 = \frac{4}{3} \, \frac{E_1^3}{c^3} \, D_D^2.
\ee
This approximation for $\gamma_1$ given by Eq.~(\ref{Einstein}) 
neglects spectral shifts and broadening.
Thus our expression for $R_0$ becomes
\be\label{R0-2}
R_0^6 &=& \frac{3}{16} \, (1 - r^2) \kappa^2 \, \left(\frac{c}{E_1}\right)^3\\\
\nonumber
& & \frac{(\gamma_2 + \gamma)  D_A^2}{(E_1  - E_2 - E_3)^2
+ (\gamma_1+\gamma_2 + \gamma)^2/16}~. \,
\ee

\medskip
Comparing Eq.~(\ref{R0-2}) and Eq.~(\ref{R0:stan}),
we see that our result differs from the standard
one in several ways. In particular, the width of the Lorentzian is 
$(\gamma_1+\gamma_2 +\gamma)/4$ instead of $\gamma/2$, 
and the factor
$(E_2/E_1)^2$ is absent in  Eq.~(\ref{R0-2}).
However, for realistic models we expect that $\gamma >> \gamma_1, \gamma_2$, hence $r << 1$.
So when we compare the values at the resonant energy
(where $E_1  - E_2 - E_3 = 0$) and if we set $Q_D=1$, 
then we get the ratio
\be
\frac{R_0}{R_F} = 0.7783 \left(\frac{E_1}{E_2}\right)^{1/3}~.
\ee
We expect that
$E_3 \simeq E_2/10$ (the energy difference between the donor's emission peak and the
acceptor's absorption peak), and thus $E_1/E_2 \simeq 1.11$, 
so we find a quite close agreement with the standard result.

\section{Application: quantum dots and bacteriorhodopsin}
Recent proposals for improved dye-sensitized solar cells \cite{Grat2003,Eijt_APL_2009} 
involve replacing the liquid dye by 
nanoparticles attached to a substrate and exploiting
FRET to achieve efficient energy transfer \cite{HHM2010}.
One candidate material is a  mixture of QDs and the protein
bacteriorhodopsin (bR) 
\cite{LLBB2007}, \cite{RRND2009}, \cite{Gri-Fr2010}. 
In one scenario the QD would act as an antenna for photon
absorption, with subsequent transfer to the retinal complex in bR.
The retinal complex in bR is known to be an efficient absorber of photons through direct
capture, and this same efficiency is expected for non-radiative transfer of excitons
from QD to bR via the FRET mechanism.\cite{BMG2010}
The methods developed in this paper can be used to evaluate the efficiency of FRET in this
hybrid system.
The QD has a band gap of approximately $2$ eV
(depending on its diameter),\cite{weber2002,crooker2002} and
after photon absorption it rapidly relaxes to 
its lowest energy excited state,
thus the QD is well modeled as a two-state system.

In its ground state the retinal molecule 
has a planar conformation. Upon excitation
it briefly enters a band of planar excited states 
(due to an electronic transition consistent with the Franck-Condon principle), and then rapidly relaxes to 
a non planar conformation.\cite{bR_coherent} 
The latter transition occurs within 
a few hundred femtoseconds, is effectively irreversible,
and thus signals the transfer of the excitation to bR.
The planar excited state lies in a band of closely spaced levels corresponding to different vibrational and
rotational states. Recent studies have demonstrated that coherence persists in the exciton state
for several hundred femtoseconds after initial excitation \cite{bR_coherent}.

\begin{figure}
\begin{center}
\includegraphics[angle=0,width=0.8 \linewidth]{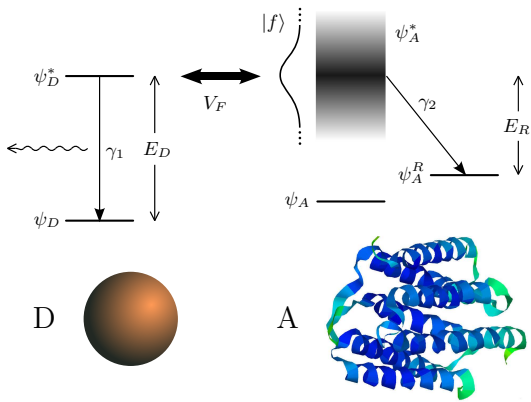}
\caption{(Color online) Schematics of the continuum model
for FRET. The donor is a quantum dot while the acceptor is an opto-electric protein
such as bacteriorhodopsin.}
\end{center}
\end{figure}
Our model QD/bR is schematized in Fig.~1 and it is a simplified
version of the more general model described above.
The values of the various parameters can be obtained from known properties of the 
molecules. The wave function $f$ has the Lorentzian form, centered at the
exciton energy $E_3$. We assume transfer on resonance, so that $E_1=E_2 + E_3 = 2$ eV. 
The width $\gamma$ is the inverse lifetime of the exciton, which is known from coherence analysis to be at least
$100$ fs, so $\gamma$ is upper bounded by around $0.05$ eV.
The rate $\gamma_1$ is set by the QD 
experimental lifetime,\cite{crooker2002} 
so $\gamma_1^{-1} = 16$ ns, and 
the rate $\gamma_2^{-1} = 500$ fs.\cite{bR_coherent} 
However, they are not important in the calculations
since they are much smaller than $\gamma$.
The FRET coupling strength $U$ 
is determined by the formula \cite{LRNB2003}
\be
U = \frac{\kappa \, D_{QD} \, D_{bR}}{\epsilon_r \, R^3}, \, 
\ee
where $\epsilon_r$ is the permittivity of the medium, $R$ is the separation
between the molecules, $D_{QD}$ and $D_{bR}$ are the dipole moments of the QD and bR respectively,
and the angular factor $\kappa$ depends on the orientations of dipoles relative
to the separation between molecules.
We use values $D_{QD}= D_{bR} = 10$ D, $\epsilon_r = 1$ (permittivity of medium, assumed dry), and
$\kappa=1$ and keep the separation distance $R$ as a free parameter. 
In atomic units this gives
\be
U = \frac{15.479}{R^3}.
\ee
\begin{figure}
\begin{center}
\includegraphics[angle=0,width=0.8 \linewidth]{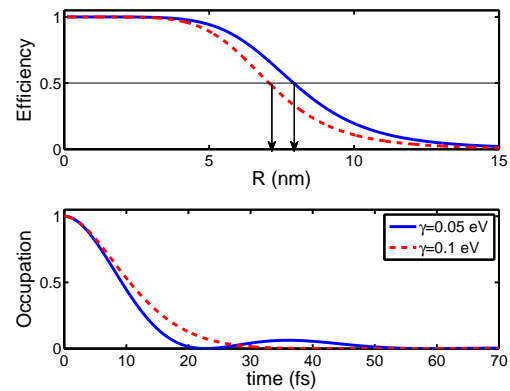}
\caption{(Color online) (a) Efficiency for QD-bR as a
function of separation; (b) occupation probability of initial excited
state as a function of time. Arrows show the position of $R_0$ for
two values of $\gamma$. }
\end{center}
\end{figure}
Figure 2 (a) shows the efficiency as a function of $R$ for these values. 
The FRET distance $R_0$ is consistent with the one estimated from experiments
of about $7-8$ nm \cite{Gri-Fr2010}. The curve almost exactly matches
the phenomenological formula \cite{Clegg,Scholes2003,For1946} for efficiency
$E = R_0^6/(R_0^6 + R^6)$. Figure 2 (b) shows the occupation probability of the initial donor 
excited state as a function of time, for the same parameter values and 
with a separation $R=2$ nm. Coherent oscillations are apparent if $\gamma=0.05$ eV.

\section{Conclusions}
We have introduced an exactly solvable model for FRET, which captures the key features of F\"orster's 
electronic-excitonic coupling in a microscopic quantum mechanical setting.
The standard F\"orster equations are accurate when the following conditions are satisfied:\cite{Scholes2003}
\begin{itemize}
\item[a)] The dipole-dipole approximation for the electronic coupling can be employed appropriately for the donor-acceptor interaction.
\item[b)] Interactions among donors or acceptors and static disorder effects leading to spectral line broadening can be neglected.
\item[c)] The energy transfer dynamics is incoherent.
\end{itemize}
Our approach goes at least beyond condition c since,
at short distances and times,
the formalism is able to capture the coherent energy transfer. 
The model uses a master equation with Lindblad operators to take account of
fluorescence and relaxation effects, and uses a continuum of excited states in the acceptor to implement
the exciton dynamics. The model is robust and can easily be extended to include more complicated exciton dynamics. As a concrete application the
model is used to analyze FRET coupling between a QD and bR, 
where it makes realistic predictions of the FRET distance.

\acknowledgments
We are grateful to Nicolas Bouchonville, 
Michael Molinari and Paul Champion for useful discussions.
B.B. was supported by US Department
of Energy, Office of Science, Basic Energy Sciences Contract 
Nos. DE-FG02 07ER46352 and DE-FG02-08ER46540 (CMSN)
and benefited from the allocation of supercomputer time at
NERSC and Northeastern University's Advanced Scientific Computation Center (ASCC). V.R. was supported by the NSF, the Wallace H. Coulter
Foundation, USAFOSR, ONR, NIH, and Harvard Medical School.
V.R. acknowledges the Rothschild Foundation 
and Varun for providing support.

\appendix

\section{}
We use resolvent techniques to compute the exponential of $B$.
Recall the resolvent representation
\be\label{int-eqn2}
e^{- i B s} = \frac{1}{2 \pi i} \, \oint \, e^{- i z s} \, (z - B)^{-1} \, d z,
\ee
where the line integral encloses the spectrum of $B$ in the complex plane. 
The assumption that $\gamma_1, \gamma_2 \ge 0$
implies that $B$ has no spectrum in the upper half-plane, so the resolvent
$(z - B)^{-1}$ is analytic in the upper half-plane. There is a cut along the real axis where the spectrum
of $h_2$ lies, and also, possibly, poles in the lower half-plane. The line integral encloses the cut and also any poles
in the lower half-plane. In the lower half-plane the integration contour can be deformed to a 
large semicircle $z = R e^{- i \theta}$ 
with $0 \le \theta \le \pi$. For $s > 0$ the contribution of this semicircle vanishes in the limit $R \rightarrow \infty$,
thus for $s > 0$ the line integral in Eq.~(\ref{int-eqn2}) 
can be written as
\be\label{int-eqn2a}
e^{- i B s} &=& - \frac{1}{2 \pi i} \, \int_{-\infty + i \epsilon}^{\infty + i \epsilon} \, e^{- i z s} \, (z - B)^{-1} \, d z\\
\nonumber
&=& \frac{1}{2 \pi i} \, \int_{-\infty + i \epsilon}^{\infty + i \epsilon} \, e^{- i z s} \, (B - z)^{-1} \, d z.
\ee
This leads to the formula
\be\label{int-eqn3a}
\bra \psi_0 | e^{- i B s} | \psi_0 \ket 
&=& \frac{1}{2 \pi i} \, \int_{-\infty + i \epsilon}^{\infty + i \epsilon} \, e^{- i z s} \, \\
\nonumber
&&\bra \psi_0 | (B - z)^{-1} | \psi_0 \ket  \, d z.
\ee
We next derive an explicit formula for the matrix element 
$\bra \psi_0 | (B - z)^{-1} | \psi_0 \ket$ appearing on the 
right-hand side above, under the assumption that ${\rm Im} z > 0$.
Define
\be
(B - z)^{-1}=
\bmx  I_{11}(z) & I_{12}(z) \cr I_{21}(z) & I_{22}(z)  \emx~.
\ee
Then the Feshbach method yields $I_{11}(z)$,
\be\label{resol7a}
\bra \psi_0 | (B - z)^{-1} | \psi_0 \ket &=& \\ 
\nonumber 
&=& \left(E_1  - \frac{i}{2} \, \gamma_1 - z - |U|^2 \,M  \right)^{-1},
\ee
where
\be
M=\bra f | (E_2 + h - \frac{i}{2} \, \gamma_2 -z)^{-1} | f \ket~.
\ee
Using the Lorentzian form for $f$, and the diagonal energy operator $h$, the matrix element $M$ is
\be
M  &= & \frac{\gamma}{2 \pi} \, \int_{-\infty}^{\infty} \frac{1}{k  - E_3 + \frac{i}{2} \gamma} \,\\
\nonumber 
&&\frac{1}{k - E_3 - \frac{i}{2} \gamma} \, \frac{1}{E_2 + k  - \frac{i}{2} \, \gamma_2 - z} \, d k.
\ee
This integral may be computed by completing the contour in the lower 
half-plane and evaluating the sum of the residues.
For ${\rm Im} z > 0$ this gives
\be
M = \frac{1}{E_2  + E_3 - \frac{i}{2} \, (\gamma +\gamma_2) - z}.
\ee
Inserting this into Eq.~(\ref{resol7a}) leads to the expression
\be\label{resol7b}
I_{11}(z) &= & (E_1  - \frac{i}{2} \, \gamma_1 - z \\
\nonumber
&& -  \frac{|U|^2}{E_2  + E_3 - \frac{i}{2} \, (\gamma + \gamma_2) - z})^{-1}
\ee
\medskip
The key observation now is that Eq.~(\ref{resol7b}) is the resolvent of the reduced two-state system
defined by the matrix introduced in Eq.~(\ref{def:B3}); that is,
\be
{\hat B} =   \bmx E_1  - \frac{i}{2} \, \gamma_1 & \overline{U} \cr
 U  &  E_2  + E_3 - \frac{i}{2} \, (\gamma + \gamma_2)  \emx~.
\ee
It follows that
\be
\bra \psi_0 | (B - z)^{-1} | \psi_0 \ket = \bra \psi_0 | ({\hat B} - z)^{-1} | \psi_0 \ket~,
\ee
and hence we obtain for all $s$
\be
\bra \psi_0 | e^{- i B s} | \psi_0 \ket = \bra \psi_0 | e^{- i {\hat B} s} | \psi_0 \ket~. 
\ee

\section{}
The acceptor's molar absorption coefficient
can be computed by using the following time-dependent
Hamiltonian for the electronic transition in the presence of
a classical field:
\be
H(t) = \bmx E_2 & \nu e^{- i \omega t} \cr \overline{\nu} e^{i \omega t} & 0  \emx~.
\ee
Here $E_2$ is the energy of the excited state, 
$\omega$ is the frequency of the radiation, and
$\nu$ is the coupling between the field and the system. 
This coupling is given by
\be\label{def:lambda}
\nu = \frac{i e}{m c} \int \overline{\psi}_{exc}(r)
\left( {\bf A_0} \cdot {\bf \nabla} \right) \psi_{gr}(r) \, d^3 r,
\ee
where ${\psi}_{exc}$ and ${\psi}_{gr}$ are the excited 
and ground state wave functions, and
${\bf A_0}$ is the field strength (assumed constant).
We have used the rotating wave approximation and dropped 
the counter-rotating term proportional
to $e^{ i \omega t}$. Standard dipole approximations lead to
\be
\nu = - \frac{i}{c} \, | {\bf A_0} | \, E_2 \, D_A~.
\ee
The radiation intensity (power per unit area)
is related to the field strength through the time-averaged 
Poynting vector, and this gives
\be
I_{in} = \frac{\omega^2 | {\bf A_0} |^2}{2 \pi c}~.
\ee
Combining the electronic transition rate with the exciton amplitude,
Fermi's Golden Rule gives the following rate for transitions in this radiation field:
\be
\sigma = 2 \pi \, |\nu|^2 \, |f(\omega  - E_2)|^2
\ee
The absorption coefficient determines the rate of energy absorption by an ensemble of molecules.
Consider a slab of absorber with unit cross-sectional area,
and thickness $x$, illuminated by light of frequency $\omega$ and intensity $I_{in}$. Then the 
output intensity is given by the Beer-Lambert law
\be\label{B-L-Law}
I_{out} = I_{in} 10^{- \epsilon_A k x},
\ee
where $\epsilon_A$ is the molar absorption coefficient and $k$ is the concentration (in moles per unit volume).
The number of molecules in the slab is $N_A k x$, 
where $N_A$ is Avogadro's number.
The energy absorbed per unit time by transitions is thus $N_A k x \, \sigma \,\omega$. 
Equating this to the energy difference between input and output gives 
the result in atomic units:
\bee
\epsilon_A \ln(10) &=& \frac{N_A \, \sigma \, \omega}{I_{in}}  \\
&=&
\frac{N_A \omega 2 \pi \, |\nu|^2 \, |f(\omega  - E_2)|^2}{I_{in}} \\
&=&
\frac{2 \pi}{\omega c}  \, \frac{N_A E_2^2 D_A^2 \gamma}{(\omega  - E_2 - E_3)^2
+ \gamma^2/4}~.
\eee


\begin{thebibliography}{99}
\bibitem{foot1}
 Electronic excitation transfer (EET) is another term used in 
the literature to describe the transfer
 of energy via non radiative dipole coupling.
\bibitem{foot2}
 The only information about
 the fluorescence spectrum needed in the present model is the rate
 of fluorescence given by the jump operator. This rate can be 
 obtained from experiments.
\end{thebibliography}
\end{document}